\documentclass[aps, prl,reprint,superscriptaddress,showpacs, citeautoscript]{revtex4-1}
\usepackage{graphicx}
\usepackage{dcolumn}
\usepackage{bm}
\usepackage{natbib}
\usepackage{amsfonts}
\usepackage{amsmath}
\usepackage{amssymb}%

\begin{document}

\preprint{APS/123-QED}

\title{Direct observation of bulk charge modulations in optimally doped Bi$_{1.5}$Pb$_{0.6}$Sr$_{1.54}$CaCu$_{2}$O$_{8+\delta}$ }

\author{M. Hashimoto}
\affiliation{Stanford Synchrotron Radiation Lightsource, SLAC National Accelerator Laboratory, 2575, Sand Hill Road, Menlo Park, California 94025, USA}
\author{G. Ghiringhelli}
\affiliation{CNR-SPIN, CNISM and Dipartimento di Fisica, Politecnico di Milano, Piazza Leonardo da Vinci 32, I-20133 Milano, Italy}
\author{W.-S. Lee}
\affiliation{Stanford Institute for Materials and Energy Sciences, SLAC National Accelerator Laboratory, 2575 Sand Hill Road, Menlo Park, California 94025, USA}
\author{G. Dellea}
\affiliation{CNR-SPIN, CNISM and Dipartimento di Fisica, Politecnico di Milano, Piazza Leonardo da Vinci 32, I-20133 Milano, Italy}
\author{A. Amorese}
\affiliation{CNR-SPIN, CNISM and Dipartimento di Fisica, Politecnico di Milano, Piazza Leonardo da Vinci 32, I-20133 Milano, Italy}
\author{C. Mazzoli}
\affiliation{CNR-SPIN, CNISM and Dipartimento di Fisica, Politecnico di Milano, Piazza Leonardo da Vinci 32, I-20133 Milano, Italy}
\author{K. Kummer}
\affiliation{European Synchrotron Radiation Facility (ESRF), BP 220, F-38043 Grenoble Cedex, France}
\author{N. B. Brookes}
\affiliation{European Synchrotron Radiation Facility (ESRF), BP 220, F-38043 Grenoble Cedex, France}
\author{B. Moritz}
\affiliation{Stanford Institute for Materials and Energy Sciences, SLAC National Accelerator Laboratory, 2575 Sand Hill Road, Menlo Park, California 94025, USA}
\author{Y. Yoshida}
\affiliation{Nanoelectronics Research Institute, AIST, Ibaraki 305-8568, Japan}
\author{H. Eisaki}
\affiliation{Nanoelectronics Research Institute, AIST, Ibaraki 305-8568, Japan}
\author{Z. Hussain}
\affiliation{Advanced Light Source, Lawrence Berkeley National Lab, Berkeley, California 94720, USA}
\author{T. P. Devereaux}
\affiliation{Stanford Institute for Materials and Energy Sciences, SLAC National Accelerator Laboratory, 2575 Sand Hill Road, Menlo Park, California 94025, USA}
\author{Z.-X. Shen}
\affiliation{Stanford Institute for Materials and Energy Sciences, SLAC National Accelerator Laboratory, 2575 Sand Hill Road, Menlo Park, California 94025, USA}
\affiliation{Geballe Laboratory for Advanced Materials, Departments of Physics and Applied Physics, Stanford University, California 94305, USA}
\author{L. Braicovich}
\affiliation{CNR-SPIN, CNISM and Dipartimento di Fisica, Politecnico di Milano, Piazza Leonardo da Vinci 32, I-20133 Milano, Italy}

\date{\today}

\begin{abstract}
Bulk charge density modulations, recently observed in high critical-temperature ($T_\mathrm{c}$) cuprate superconductors, coexist with the so-called pseudogap and compete with superconductivity. However, its direct observation has been limited to a narrow doping region in the underdoped regime. Using energy-resolved resonant x-ray scattering we have found evidence for such bulk charge modulations, or soft collective charge modes (soft CCMs), in optimally doped Bi$_{1.5}$Pb$_{0.6}$Sr$_{1.54}$CaCu$_{2}$O$_{8+\delta}$ (Pb-Bi2212) around the summit of the superconducting dome with momentum transfer $q_{\parallel}\sim0.28$ reciprocal lattice units (r.l.u.) along the Cu-O bond direction. The signal is stronger at $T\simeq T_\mathrm{c}$ than at lower temperatures, thereby confirming a competition between soft CCMs and superconductivity. These results demonstrate that soft CCMs are not constrained to the underdoped regime, suggesting that soft CCMs appear across a large part of the phase diagram of cuprates and are intimately entangled with high-$T_\mathrm{c}$ superconductivity.
\end{abstract}


\maketitle
 A much studied property of high critical-temperature ($T_\mathrm{c}$) cuprate superconductors is the opening, below the temperature $T^*$ ($> T_\mathrm{c}$), of the so-called \emph{pseudogap} \cite{Timus99}. Recent experiments have suggested that the pseudogap is accompanied by a special broken electronic symmetry, often indicated as a modulation of the charge density residing in the CuO$_2$ planes \cite{Tranq95,Tranq04,Abbam05,Hucker11,Wilki11,Dean13a,Thamp13,Fink11,Ghiri12,Chang12,Achka12,Thamp13,Black13,Tacon,SilvaNeto13,Comin13,Doiro07,Sebas08,Hoffm02,Howal03,Versh04,Kohsa07,Kohsa08,
Fauqu06,Li08,Hashi10,Taill10,Wu11, He11,Shekh13,Torch13}. These charge modulations can be seen as soft charge collective modes (soft CCMs), comprising both static charge density waves with no energy loss, and dynamical charge modes having finite mass. If the pseudogap was connected to a broken translational symmetry, the soft CCMs would be truly static. Soft CCMs in bulk states have been observed only in a narrow interval of the underdoped regime, there establishing a clear link with the $T_\mathrm{c}$ suppression around 1/8 doping (``the 1/8 anomaly''). On the other hand, causality links with the opening of the pseudogap still need to be demonstrated. How soft CCMs shape the Fermi surface and how strongly they couple to the phonon dispersion is still actively studied. Further, knowing their doping dependence and how they evolve from static charge order to dynamical charge modes may provide crucial insights about their role in high-$T_\mathrm{c}$ superconductivity and about the quantum critical point in the cuprate phase diagram.

Evidence for soft CCMs in the bulk states around $p = 1/8$ hole doping was provided first indirectly by inelastic neutron scattering (INS), as a magnetic incommensurate peak near the antiferromagnetic points $(0.5\pm \simeq 0.125, 0.5, L)$ and $(0.5, 0.5\pm \simeq 0.125, L)$ in La$_{1.6-x}$Nd$_{0.4}$Sr$_x$CuO$_4$ (LNSCO) and La$_{2-x}$Ba$_x$CuO$_4$ (LBCO) \cite{Tranq95,Tranq04}. Later, the corresponding charge peak was observed by x-ray scattering at the incommensurate wavevector $q_{\parallel}\simeq 0.25$ along the Cu-O bond direction in LBCO, LNSCO \cite{Abbam05,Hucker11,Wilki11,Dean13a,Thamp13}, and La$_{1.8-x}$Eu$_{0.2}$Sr$_x$CuO$_4$ (LESCO) \cite{Fink11}. Only more recently, soft CCMs, $q_{\parallel}\simeq 0.3$ reciprocal lattice units (r.l.u.), have been observed around $p=1/8$, in YBa$_2$Cu$_3$O$_{6+\delta}$ (YBCO) and NdBa$_2$Cu$_3$O$_{6+\delta}$ (NBCO) \cite{Ghiri12,Chang12,Achka12,Thamp13,Black13,Tacon}, and in Bi$_2$Sr$_2$CaCu$_2$O$_{8+\delta}$ (Bi2212) \cite{SilvaNeto13} and Bi$_{1.5}$Pb$_{0.55}$Sr$_{1.6}$La$_{0.4}$CuO$_{6+\delta}$ (Bi2201) \cite{Comin13}, respectively. Particularly, in YBCO and Bi2212, the temperature dependence of the soft CCM scattering peak indicates that they are in competition with superconductivity.
On the other hand, evidence for translational symmetry-breaking has been accumulating indirectly via other techniques. Under magnetic fields, in YBCO, Fermi surface reconstruction has been reported by quantum oscillation studies \cite{Doiro07,Sebas08} and evidence for a charge order has been found by nuclear magnetic resonance (NMR) \cite{Wu11}. Scanning tunneling microscopy (STM) has shown complex charge modulations of the surface electronic states in Bi2201, Bi2212 and Ca$_{1.88}$Na$_{0.12}$CuO$_2$Cl$_2$ (CNCOC) \cite{Hoffm02,Howal03,Versh04,Kohsa07,Kohsa08,SilvaNeto13,Comin13}. Angle-resolved photoemission (ARPES) studies also have suggested that the dispersion in the antinodal region is consistent with translational symmetry-breaking \cite{Hashi10, He11}. These charge modulations have been observed in a wide doping range, but questions arise about the correspondence of bulk electronic states with those studied at the surface of samples.

In this Rapid Communication, we have studied optimally doped Bi$_{1.5}$Pb$_{0.6}$Sr$_{1.54}$CaCu$_{2}$O$_{8+\delta}$ (Pb-Bi2212) by high resolution resonant inelastic x-ray scattering (RIXS), focusing on the quasi-elastic spectral component that is sensitive to charge modulations \cite{Ghiri12}. We have found a peak at $q_{\parallel}\simeq0.28$ r.l.u. along the Cu-O bond direction, indicating  the presence of bulk, incommensurate, soft CCMs in the pseudogap state. The associated peak is weaker at low temperatures, in the superconducting state, than at $T_\mathrm{c} = 98$K. The direct observation of soft CCMs in an optimally doped compound significantly expands the doping range where superconductivity, the pseudogap, and the tendency towards charge ordering coexist, hinting at the universality of soft CCMs in the cuprate superconductors and thus stimulating different hypotheses about their role in superconductivity itself.

The high quality optimally doped Pb-Bi2212 single crystal was grown by the floating zone method. The hole concentration was optimized by annealing the samples in N$_2$ flow. The RIXS measurements were performed with the AXES spectrometer at the beamline ID08 of the European Synchrotron Radiation Facility (ESRF) \cite{Dallera96}. We used the incident photon energy of $\sim$931 eV, at the maximum of the Cu $L_3$ absorption peak (2$p_{3/2} \to 3d$ transition). The combined energy resolution was $\sim$0.26 eV, and the illuminated area on the sample surface was 7 $\mu$m $\times$ 60 $\mu$m. The sample size was approximately $2 \times 2 \times 0.5$ mm$^3$. The sample was measured twice, and for each experimental run it was cleaved in air some minutes before installation inside the ultra-high vacuum measurement chamber ($\sim 3 \times 10^{-9}$ mbar). The sample temperature was varied between 20 K and 96 K. We have checked that incommensurate structural modulations of the BiO planes do not affect the RIXS spectra along (100), i.e., the Cu-O bond direction.

\begin{figure}
\includegraphics[width=\linewidth]{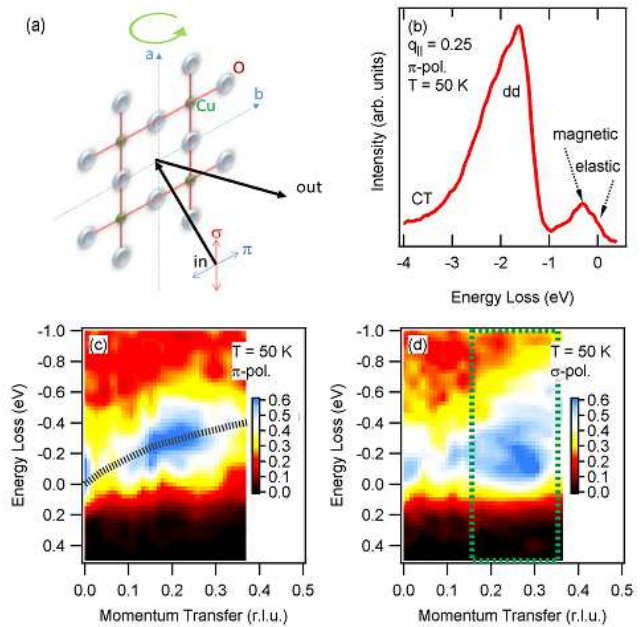}
\caption{Resonant soft x-ray scattering from optimally doped Bi2212. (a) Scattering geometry with the $c$ and either $a$ or $b$ axes in the scattering plane. The incident photon polarization can be parallel ($\pi$) or perpendicular ($\sigma$) to the scattering plane. (b) RIXS spectrum at $q_{\parallel}\simeq$ 0.25 r.l.u. Quasi-elastic peak, phonon excitations ($\sim$-0.07 eV), dispersing paramagnons ($<$-0.35 eV), and the $d$-$d$ excitations around -2 eV are indicated. (c) and (d) Low-energy RIXS intensity map taken at $T\sim$50 K along the Cu-O bond direction with $\pi$ and $\sigma$ polarizations, respectively. The temperature dependent measurements, shown in the next figures, were carried out with $\sigma$ polarization in the momentum transfer region indicated by the box.}
\label{Fig1}
\end{figure}

We show in Fig. 1(a) the experimental geometry. The scattering angle between the incident and outgoing photons was fixed at 130 deg. The in-plane momentum transfer along the Cu-O bond direction was scanned by rotating the sample. We used both $\pi$ and $\sigma$ polarizations, which are parallel and perpendicular to the scattering plane, respectively. Note that the out-of-plane momentum transfer along the $c$-axis ($q_{\perp }$) also changes with sample rotation; however, it has been shown that the scattering signals for the CCMs in the cuprates are nearly independent of $q_{\perp }$ \cite{Abbam05, Ghiri12, Comin13, SilvaNeto13}. A typical RIXS spectrum, shown in Fig. 1(b) for $q_{\parallel}$=0.25 r.l.u., consists of $d$-$d$ and charge transfer (CT) excitations at higher energy, magnetic and phonon excitations and the elastic peak at low energies \cite{Ament11}. We show in Figs. 1(c) and 1(d) the intensity maps at 50 K ($<T_\mathrm{c}$) for q$_{\parallel} \ge0$ r.l.u. along the Cu-O bond direction, with $\pi$ and $\sigma$ polarizations, respectively. All spectra in all figures are normalized to the total intensity (integrated over the -3.5 eV and 1.0 eV energy loss interval).

The RIXS spectra of Fig. 1(c), measured with $\pi$ polarization, are dominated by single spin-flip excitations, particularly at large $q_{\parallel}$ values \cite{Braic10a, Sala11}. The dashed line highlights the dispersion of a broad spectral feature that reaches the maximum energy of $\sim$-350 meV at the zone boundary. These are paramagnon excitations, already observed by RIXS \cite{Braic10,LeTa11, Dean13b, LeTa13}. On the other hand, the map of Fig. 1(d), measured with $\sigma$ polarization, is composed of non-spin-flip and double spin-flip processes (elastic scattering, charge excitations, phonons and bi-magnons)\cite{Braic10a, Sala11}. In this work, we focus on the low-energy excitations with $\sigma$ polarization, particularly around $q_{\parallel} \sim$0.25 r.l.u., where the quasi-elastic signal reaches a maximum intensity. This type of signal, although weak, is similar to the signature of soft CCMs detected in RIXS spectra of YBCO by Ghiringhelli $et al$. \cite{Ghiri12}. To detect broad and weak modulations of the quasi-elastic signal in Bi2212, the energy resolution of RIXS is crucial. With conventional scattering measurements that integrate over the whole inelastic spectrum shown in Fig. 1(a), the tiny quasi-elastic signal would be easily lost, buried under the large inelastic signal \cite{Ghiri12}.

\begin{figure}
\includegraphics[width=\linewidth]{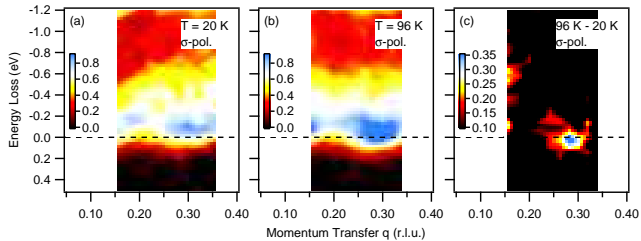}
\caption{Temperature dependence of the RIXS spectra across $T_\mathrm{c}$. (a) and (b) RIXS intensity map taken with $\sigma$ polarization at $T\sim$20 K ($\sim T_\mathrm{c}$) and 96 K ($\ll T_\mathrm{c}$), respectively. (c) Color map for 96 K -- 20 K showing a strong temperature dependence at $q_{\parallel}\sim$0.28 r.l.u.}
\label{Fig2}
\end{figure}

We show in Figs. 2(a) and 2(b) the color intensity maps of the RIXS spectra at $T = 96$ K ($\sim T_\mathrm{c}$) and $T = 20$K ($\ll T_\mathrm{c}$), respectively. At both temperatures the low energy part of the spectra is stronger at $q_{\parallel} \simeq 0.28$ r.l.u., but the peak appears more intense at $T \simeq T_c$, as made evident by the map of Fig. 2(c) obtained by subtracting the low-$T$ from the high-$T$ data sets. In the difference map the peak at $q_{\parallel} \simeq 0.28$ r.l.u. is very close to the zero energy loss position. However, it is not possible to conclude whether the peak is exactly at zero energy loss due to experimental energy resolution.

\begin{figure}
\includegraphics[width=\linewidth]{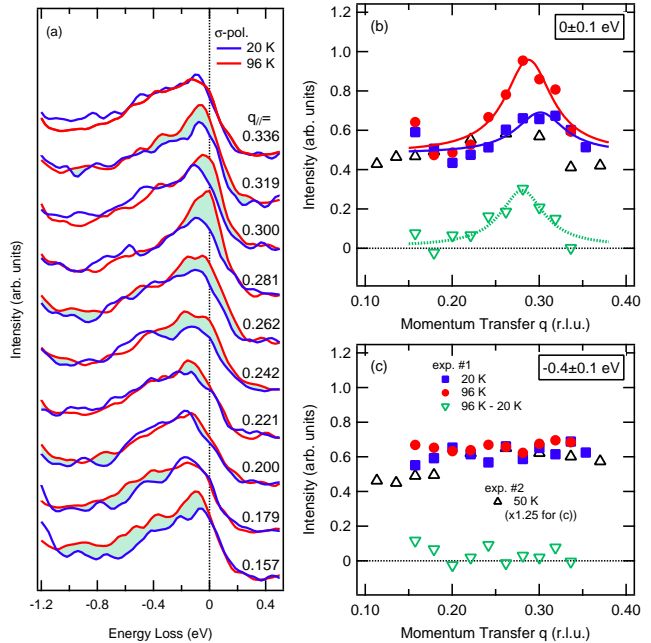}
\caption{A signature of the soft CCMs at $q_{\parallel}\sim$0.28 r.l.u. (a) RIXS spectra at 96 K (red) and 20 K (blue) at selected $q_{\parallel}$. The differences between the two temperatures are highlighted by the green shading. (b) Temperature dependence of the intensity at 0$\pm$0.1 eV for the quasi-elastic signal and (c) -0.4$\pm$0.1 eV for the multi-magnon signal, respectively. The 20 K and 96 K intensities are from the experiment No.1 and the 50 K intensities are from the experiment No.2. The differences between 96 K and 20 K are also plotted (open green triangles).}
\label{Fig3}
\end{figure}

To evaluate this temperature dependence more quantitatively, we show in Fig. 3(a) the $\mathbf{q}$ dependence of the RIXS spectra at low energies at $T\sim T_\mathrm{c}$ (red) and $T \ll T_\mathrm{c}$ (blue). Differences between the two temperatures are highlighted by green shading. This representation confirms the effect close to zero energy loss at $q_{\parallel}\simeq0.28$ r.l.u. This is seen more clearly in the constant-energy cuts of Fig. 3(b), obtained by integrating the intensity over a 200 meV band centered at zero energy loss at 20 K and 96 K in red circles and blue squares, respectively. We can thus evaluate the full width at half maximum (FWHM) of the peak at $q_{\parallel} \simeq0.28$ r.l.u. to be $>$0.05 r.l.u. at $T_c$. This peak becomes significantly weaker and possibly broader at lower temperatures but remains visible down to 20K. Also in this representation, the difference between 96 K and 20 K [open green triangles in Fig. 3(b)] highlights the peak at $q_{\parallel}\sim 0.28$ r.l.u. As a reference, we report in Fig. 3(c) the corresponding -0.4$\pm$0.1 eV constant-energy cuts. At this energy spectra are dominated by the multi-magnon excitations and the result shows an almost negligible momentum and temperature dependence. We note that the quasi-elastic intensities at 50 K from experiment No.2 [black triangles in Fig. 3(b)] are weaker than those for experiment No.1, although the enhancement of the quasi-elastic signal at $q_{\parallel} \simeq 0.28$ r.l.u. is observed consistently. This is probably due to differences in surface quality. To better compare the intensity profiles along $q_{\parallel}$ the intensity at 50 K from experiment No.2 is multiplied by a factor of 1.25.

The peak of the quasi-elastic signal at $q_{\parallel}\simeq0.28$ r.l.u. suggests the existence of soft CCMs with a periodicity of $\sim 3.57$ lattice units in the bulk state of optimally doped Bi2212. The result is very similar to previous scattering reports from underdoped samples. The only  previous observation of soft CCMs in an optimally doped sample is on LBCO, which however undergoes a structural transition not present for other cuprates. There the stabilization of static charge order is abrupt across the orthorhombic-tetragonal transition at 56 K \cite{Hucker11}, although the soft CCM signal was detected with RIXS up to 85 K \cite{Dean13a}. Therefore, these previous results (see Table I) cannot conclusively establish whether soft CCMs exist in a wider doping range including optimal doping ($\sim$16$\%$).

On the other hand, a short range checkerboard or stripe-like order over a wide doping range has been reported by STM \cite{Hoffm02,Howal03,Versh04,Kohsa07,Kohsa08, Kohsa07}. The characteristic wave vector is $q_{\parallel}\simeq 0.25$ r.l.u. along the Cu-O bond direction \cite{Hoffm02,Howal03,Versh04,Kohsa07,Kohsa08}. Further, the dispersion in the pseudogap state, determined by ARPES, is consistent with short correlation-length charge modulations \cite{Hashi10, He11}. However, STM and ARPES are surface sensitive probes, and cannot confirm the bulk electronic states. Our RIXS results on Bi2212 demonstrate that bulk charge modulations exist at least up to optimal doping, suggesting that the soft CCMs are universal in the pseudogap regime of the cuprates. Importantly, the observed characteristic wave vector of $q_{\parallel}\simeq 0.25$ and the short correlation length are, though not identical, consistent between the bulk and surface sensitive studies.

The drastic decrease of the $q_{\parallel}\simeq 0.28$ r.l.u. peak below $T_\mathrm{c}$ [Fig. 3(b)] shows that soft CCMs are suppressed by superconductivity and compete with it, in analogy to what has been observed in YBCO, where the soft CCM peak weakens and broadens below $T_\mathrm{c}$, but is restored by an external magnetic field \cite{Ghiri12,Chang12,Thamp13, Black13}. We note that at $T_\mathrm{c}$ the soft CCM peak is broader in Bi2212 than in YBCO, and that the incommensurate wave vector is slightly different. From the data of Table I we see that the soft CCM wave-vector spans the 0.24-0.31 r.l.u. range in various materials, and the peak intensity and width are influenced by compositional disorder. 
It is also noteworthy that ARPES has suggested a competition between the pseudogap and superconducting order parameters in Bi2212 \cite{Hashi}, establishing an important connection to the competition between soft CCMs and superconducting order parameters seen here. Further, such a phase competition could results in the recently proposed phase diagram \cite{Vishik12} where the phase boundary for the competing order bends back in the superconducting dome.

\begin{table}
	\caption{Comparison of charge modulations in different cuprate compounds measured with soft x-ray scattering. $\xi$ is the correlation length as deduced from the full width as half maximum $w$ of the scattering peak, $\xi = a/(\pi * w)$.}
	\begin{tabular}{lccccr}\hline
		Sample & $p$ & $T_\mathrm{c}$ & $q_{\parallel}$ (r.l.u.) & $\xi$ (\AA)               & refs. \\ \hline \hline
		Bi2201 & 0.115 & 15           & 0.265                 & 26                        & \cite{Comin13} \\
		Bi2201 & 0.130 & 22           & 0.257                 & 23                        & \cite{Comin13} \\
		Bi2201 & 0.145 & 30           & 0.243                 & 21                        & \cite{Comin13} \\
		Bi2212 & 0.09  & 45           & 0.30                  & 24                        & \cite{SilvaNeto13} \\
		Bi2212 & 0.160 & 98           & 0.28                  & $<$24 (at $T_\mathrm{c}$) & This work \\
		YBCO   & 0.115 & 61           & 0.32                  & $\sim$60 (at $T_\mathrm{c}$)& \cite{Ghiri12, Thamp13} \\
		LBCO   & 0.125 & 2.5          & 0.236                 & $\sim$200                 & \cite{Abbam05, Wilki11,Thamp13,Dean13a} \\
		LBCO   & 0.155 & 30           & 0.244                 & $\sim$240 (15 -- 25 K)    & \cite{Abbam05, Wilki11,Thamp13,Dean13a} \\ \hline

	\end{tabular}
\end{table}

Charge modulations are often driven by Fermi surface nesting, as it has been discussed also for the cuprates. However, the Fermi surface shape of optimally doped Bi2212 determined by ARPES \cite{Lee07, Kondo13, Hashi} suggests that the soft CCM wave-vector is longer than the distance between the Fermi momenta at the Brillouin zone boundary (antinodes), so that soft CCMs cannot be driven by the Fermi surface nesting between the antinodes. This is also in agreement with the non-trivial doping- and material-dependence of $q_{\parallel}$  (Table I) and the results of Ref. \onlinecite{SilvaNeto13}. The diversity of $q_{\parallel}$ and the effects of magnetic fields \cite{Doiro07, Sebas08,Wu11} call for further studies.

Although soft CCMs are not related with an obvious Fermi surface nesting vector and cannot thus directly explain the opening of the peudogap in that region of the $k$-space, a link between the two phenomena is likely to exist up to optimal doping and possibly even in the overdoped regime. These facts suggest that an incipient charge density instability is inherent to the pseudogap regime. Although our result cannot conclude whether the soft CCMs are fluctuating or static, the short correlation length in the optimally doped sample could be due to an increasingly dynamical behavior of charge modes with the doping level. The observation of the soft CCM peak gaining mass, decreasing in intensity, and increasing in width as the phase boundary is crossed would be compelling evidence for an underlying quantum critical point and the competition with superconductivity, and awaits further study with more sensitive and selective x-ray probes in the future.

\section{Acknowledgment}
This work was performed at the ID08 beam line of the ESRF (Grenoble, France). This work is supported by the Department of Energy,  Office of Science, Basic Energy Sciences, Materials Sciences and Engineering Division, under Contract No. DE-AC02-76SF00515; and by the Italian Ministry of University and Research (MIUR) through the Grants PRIN20094W2LAY and PIK-POLARIXS.

%


\end{document}